\documentclass{article}

\usepackage[authoryear]{natbib}
\usepackage{url} 
\usepackage{authblk}
\usepackage{graphicx} 
\usepackage{comment}

\begin{document}
\def\floatpagepagefraction{1}
\def\textpagefraction{.001}

\title{Nowcast of an EUV dynamic spectrum during solar flares}                      
\date{Submitted 29 November 2019; Accepted 30 April, 2020}
\author{Toshiki Kawai \thanks{Institute for Space-Earth Environmental Research, Nagoya University, Aichi, Japan } \and 
Shinsuke Imada \thanks{Institute for Space-Earth Environmental Research, Nagoya University, Aichi, Japan } \and 
Shohei Nishimoto \thanks{National Defence Academy of Japan, Kanagawa, Japan} \and 
Kyoko Watanabe \thanks{National Defence Academy of Japan, Kanagawa, Japan} \and 
Tomoko Kawate \thanks{National Institute for Fusion Science, Gifu, Japan}}

\maketitle

\begin{abstract}
In addition to X-rays, extreme ultraviolet (EUV) rays radiated from solar flares can cause serious problems, such as communication failures and satellite drag. 
Therefore, methods for forecasting EUV dynamic spectra during flares are urgently required. 
Recently, however, owing to the lack of instruments, EUV dynamic spectra have rarely been observed. 
Hence, we develop a new method that converts the soft X-ray light curve observed during large flare events into an EUV dynamic spectrum by using the {\it Solar Dynamics Observatory} / {\it Atmospheric Imaging Assembly} images, a numerical simulation, and atomic database. 
The simulation provides the solution for a coronal loop that is heated by a strong flare, and the atomic database calculates its dynamic spectrum, including X-ray and EUV irradiances. 
The coefficients needed for the conversion can be calculated by comparing the observed soft X-ray light curve with that of the simulation. 
We apply our new method to three flares that occurred in the active region 12673 on September 06, 2017. 
The results show similarities to those of the {\it Flare Irradiance Spectral Model}, and reconstruct some of the EUV peaks observed by the {\it EUV Variability Experiment} onboard the {\it Solar Dynamics Observatory}.
\end{abstract}

\section{Introduction}
A solar flare, which is a sudden brightening observed in almost all wavelengths from radio waves to gamma rays, was first observed by Richard Carrington in 1859 \citep{Carrington1859}. 
Nowadays, it is believed that Solar flare is a result of the rapid release of magnetic energy stored in the solar corona. 
The energy released by a flare is very large, often reaching $10^{32}$ ergs within an hour. 
One standard model of flares that is based on magnetic reconnection is the CSHKP model \citep{Carmichael1964, Sturrock1966, Hirayama1974, Kopp1976}. 
The characteristics predicted based on these models have been verified by modern observations ({\it e.g.}, cusp-like structure in soft X-ray images \citet{Tsuneta1992}, hard X-ray sources above the flare loop \citet{Masuda1994}, chromospheric evaporation \citet{Teriaca2003, Imada2015}, reconnection inflows \citet{Yokoyama2001}, reconnection outflows (off limb \citet{McKenzie1999, Innes2003, Imada2013}, on disc \citet{Hara2011}, plasmoid ejection \citet{Ohyama1998, Liu2013}, and coronal mass ejections \citet{Svestka1992, Imada2007}). 
X-rays and extreme ultraviolet (EUV) radiation from solar flares sometimes cause serious problems, such as radiation exposure, communication failures, and satellite drag \citep{Lean1997}. 
These incidents can make our lives inconvenient, delay space development, and cause significant economic losses.
To avoid these problems, it is urgently required that we develop the means to predict large solar flare effects on the earth.

Increases of the terrestrial upper atmosphere density can cause dangerous events for a satellite. 
Satellite drag is caused by the friction between a satellite in low earth orbit and the upper atmosphere. 
Drag increases happen when large flares occur because their X-ray and EUV emissions enhance ionization in the upper atmosphere. 
EUV irradiance, especially that emitted in the transition region and upper chromosphere, contribute significantly to ionization in the F-region ionosphere \citep{Qian2010}.
Therefore, it is important to predict not only X-ray but also EUV emissions from solar flares. 

Since 2001, the {\it Solar EUV Experiment} \citep[SEE:][]{Woods1998, Woods2005, Woods2012} onboard the {\it Thermosphere Ionosphere Mesosphere Energetics and Dynamics} (TIMED) has provided sub-daily solar EUV irradiance information from 1 to 2000 \AA. 
The {\it Solar Extreme ultraviolet Monitor} (SEM) onboard the {\it Solar and Heliospheric Observatory} (SOHO) also has measured solar irradiance from 1 to 770 \AA~ with 60 s cadence \citep{Judge1998} since 1995.
Besides this, the {\it EUV Variability Experiment} \citep[EVE:][]{Woods2012} onboard the {\it Solar Dynamics Observatory} \citep[SDO:][]{Pesnell2012} began to observe solar EUV spectrum from the full disc of the sun and the corona in 2010. 
EVE is composed of two types of {\it Multiple EUV Grating Spectrographs} (MEGS), MEGS-A and MEGS-B, which monitor the EUV spectrum from 50 to 370~\AA~and from 350 to 1050~\AA, respectively.
However, the observations of MEGS-A were terminated in 2014 due to a power anomaly, and MEGS-B also has limited daily exposure now because of an unexpected and rapid degradation. 
In addition to them, the {\it Extreme Ultraviolet Sensors} \citep[EUVS:][]{Eparvier2009} onboard the {\it Geostationary Operational Environmental Satellite R} (GOES-R) series have monitored eight lines and bands between 250 and 2850~\AA~since 2016.
From the EUVS observation, the EUVS model \citep{Thiemann2019} estimates the solar irradiance spectral between 50 and 1270~\AA~at 30 s cadence.
By contrast, solar soft X-ray irradiance has been monitored by the GOES series since 1974. 
Each GOES satellite has two X-Ray Sensors (XRS), which observe temporal emission from 0.5 to 4 \AA~(XRS-A) and from 1 to 8 \AA~(XRS-B).
To understand EUV emission during solar flares, a method of converting X-rays to EUV spectra must first be developed. 

The {\it Flare Irradiance Spectral Model} \citep[FISM:][]{Chamberlin2007, Chamberlin2008} is an empirical model that estimates the spectral emission from 1 to 1900 \AA, with a wavelength resolution of 10 \AA~and a time cadence of 60 s. 
The solar irradiance for a particular wavelength $\lambda$ at time $t$, $E(\lambda, t)$, is defined as:
\begin{equation}
E(\lambda, t) = E_{\mathrm{min}}(\lambda)+\Delta E_{\mathrm{SC}}(\lambda, t)+\Delta E_{\mathrm{SR}}(\lambda, t) \\
+\Delta E_{\mathrm{GP}}(\lambda, t)+\Delta E_{\mathrm{IP}}(\lambda, t)
\end{equation}
where $E_{\mathrm{min}}$, $\Delta E_{\mathrm{SC}}$, $\Delta E_{\mathrm{SR}}$, $\Delta E_{\mathrm{GP}}$, and $\Delta E_{\mathrm{IP}}$ represent the minimum spectral values and the variation of solar irradiance due to the solar cycle, solar rotation, and gradual and impulsive phases of the flare, respectively.
The FISM succeeded in reproducing well the irradiance variation over timescales ranging from seconds (flares) to years (cycles). 

Different from the FISM, some physical models to estimate the EUV irradiance during flares are recently developed. 
\citet{Li2012} partially succeeded in reproducing some EUV light curves emitted from a M1.0 flare by using zero-dimensional coronal loop model called the {\it Enthalpy Based Thermal Evolution of Loops} \citep[EBTEL:][]{Klimchuk2008, Cargill2012} model.
\citet{Thiemann2017} achieve the limited success in reproducing EUV light curves of some Fe ion lines by using empirical relationships under the assistance of the EBTEL model. 
However, due to lack of the calculation of the spatial distribution, it is difficult for zero-dimensional model to estimate the emissions from the transition region and lower.
These emissions are considered as the important for satellite drag problems.

\begin{figure*}
\begin{center}
\includegraphics[width=1\linewidth]{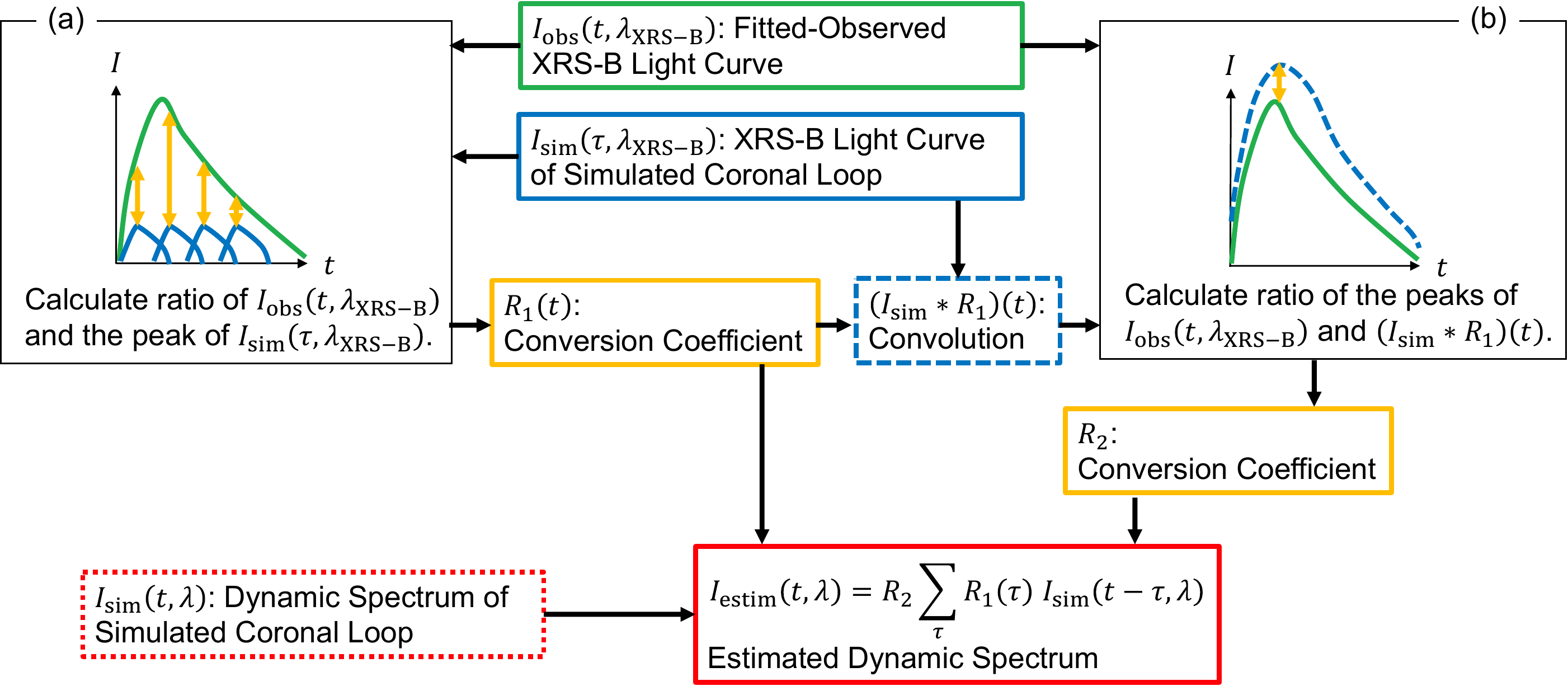}
\caption{The flow of conversion method. $R_1$ is calculated from the maximum intensity of simulated XRS-B and the observation (a). $R_2$ is calculated from the convolution of the simulated XRS-B light curve as well as $R_1$ and the observation (b). The dynamic spectrum including an EUV irradiance can be converted by applying $R_1$ and $R_2$ to equation \ref{eq:conv}}\label{fig:chart}
\end{center}
\end{figure*}

In this paper, we introduce a new method for estimating the EUV dynamic spectrum during a solar flare from the GOES X-ray light curve by using SDO/AIA images, a numerical simulation, and an atomic database. 
The simulation calculates the temporal evolution of a coronal loop heated by a flare. 
The database calculates the X-ray and EUV dynamic spectra of the simulated coronal loop. 
We estimate the conversion function of calculated XRS-B light curve to the observed one. 
Applying it to other wavelength emissions, we obtain the X-ray and EUV dynamic spectra of a solar flare. 
Figure~\ref{fig:chart} shows the flow of the conversion we describe in the following sections.

\section{Soft X-ray observation}
\begin{figure}
\begin{center}
\includegraphics[width=1\linewidth]{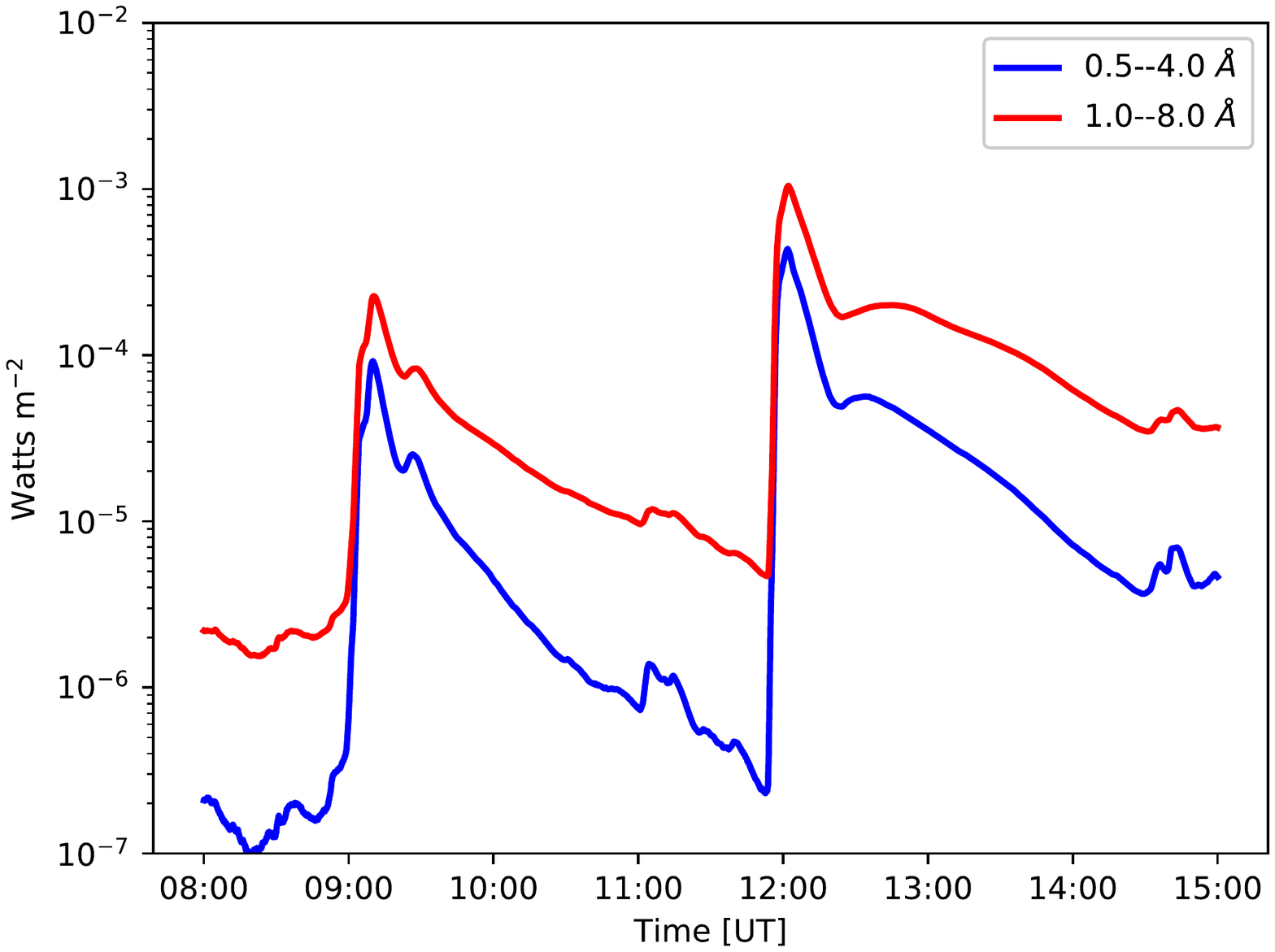}
\caption{Soft X-ray light curves observed during the flares, obtained from XRS-A (blue) and -B (red) onboard GOES13. We focus on three flares from the observations mentioned in Table~\ref{tab:flare}.}\label{fig:goes}
\end{center}
\end{figure}

We apply our method to three GOES X-class flares, which occurred in AR12673 (S09W42) between 08:50 and 14:40 UT on September 06, 2017. 
One of them is the largest flare observed in solar cycle 24, which scored X9.3. 
Figure~\ref{fig:goes} represents GOES XRS-A and XRS-B light curves during the flares. 
We use the observation of GOES13 instead of 15 because of the lack of data. 
The GOES data is available from \url{https://umbra.nascom.nasa.gov}.
We define the start, peak, and end time of each flare by using the time derivative of XRS-B. 
The times and GOES classes of each flare are given in Table~\ref{tab:flare}. 
The AR12673 and its eruptive events are reported in several papers \citep[{\it e.g.},][]{Yan2018, Inoue2018}.
Moreover, \citet{Chamberlin2018} estimates the dynamic spectra during these flares by using FISM model and compares them with the observations.

\begin{table}
\begin{center}
  \caption{The class and start, peak, and end time of each flare. Each time is obtained from the time derivative of the GOES XRS-B light curve.}
  \begin{tabular}{ccrrr}
 \hline
     & GOES class & start & peak & end \\ \hline
     Flare 1 & X2.2 & 08:57:39 & 09:10:25 & 09:23:11 \\
     Flare 2 & X9.3 & 11:53:42 & 12:02:12 & 12:24:28 \\
      Flare 3 &  X2.0 & 12:25:00 & 12:45:41 & 14:22:03 \\
       \hline
  \end{tabular}
  \label{tab:flare}
\end{center}
\end{table}

\begin{figure}
\begin{center}
\includegraphics[width=1\linewidth]{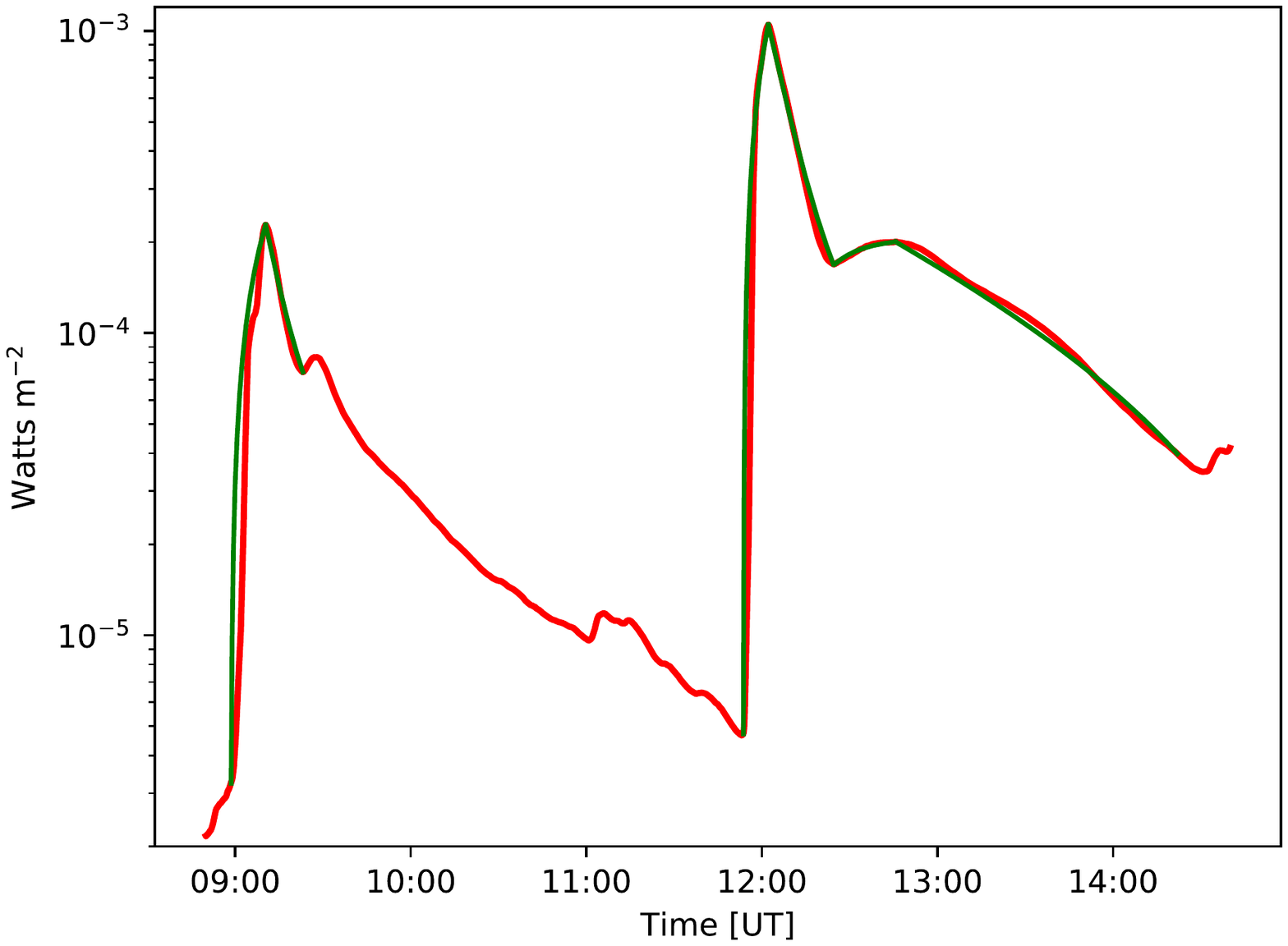}
\caption{GOES XRS-B light curve observed during the events (red) and their fitted lines (green), found using the Gauss-Newton method. }\label{fig:fit}
\end{center}
\end{figure}

\section{Method}
\subsection{Preprocessing}
To simplify the analysis, we separate the observed GOES XRS-B light curve into the rising (from start to peak) and decaying (from peak to end) phases. 
Then, we approximate the light curve of each phase using an exponential curve such as
\begin{equation}
\label{eq:tau}
I(t) = A e^{\pm{ t / \tau}} + B
\end{equation}
 where $I$ and $t$ are XRS-B intensity and elapsed time from the start (rising phase) or peak (decaying phase), respectively. 
 There are three free parameters, therefore, we set both ends of each exponential curve to match with the observed values. 
 In this case, 
\begin{eqnarray}
A = \frac{I\left(t_{e}\right) - I(0)}{e^{\pm{ t_{e} / \tau}} - 1} \\
B = I(0) - A
\end{eqnarray}
where $t_{e}$ represents the time between start and peak (rising phase) or the peak and end (decaying phase). 
We estimate the last parameter $\tau$ using the Gauss-Newton Method, which can solve non-linear least squares problems. 
We calculate errors for cases where the sign in the exponential is positive (divergent) and negative (convergent), and subsequently obtain $\tau$ with greater accuracy.
Figure~\ref{fig:fit} represents the observed and fitted light curves during the three flares. 
The obtained $\tau$ of the rising and decaying phases of each flare are given in Table~\ref{tab:tau}.

\subsection{Coronal loop length measurement}
\begin{figure*}
\begin{center}
\includegraphics[width=1\linewidth]{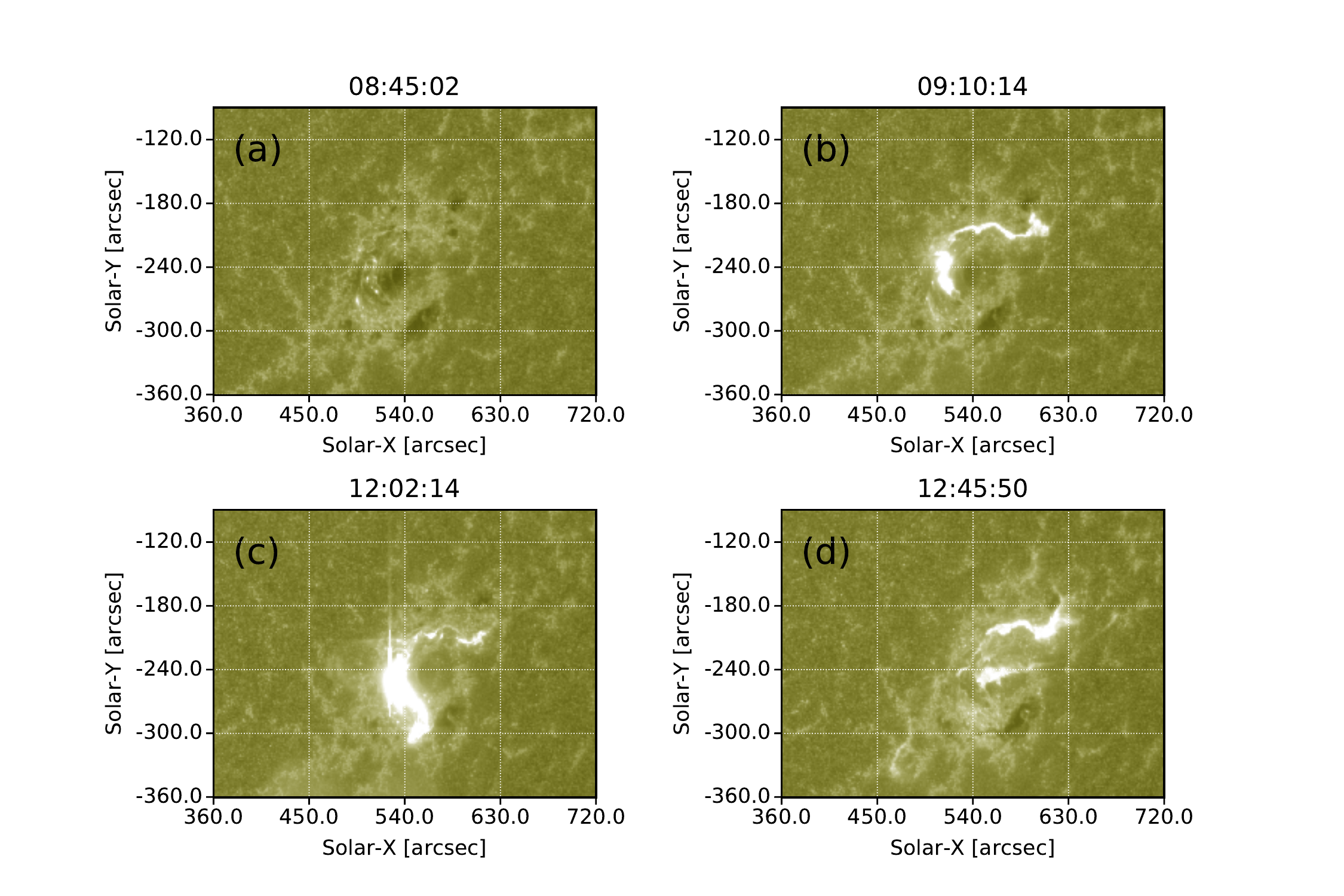}
\caption{Some snapshots obtained from the SDO/AIA 1600 \AA~ filter during the flare events. Panel (a) represents the beginning of the first flare. Some brightening points can be seen around the sunspots. Panels (b) - (d) depict the image when the GOES XRS-B light curve reaches the peak of each flare. A pair of flare ribbons exists on the active region in each panel. }\label{fig:1600}
\end{center}
\end{figure*}

For the numerical simulation, we estimate the coronal loop length that caused these three flares, using the imaging observations. 
Figure~\ref{fig:1600} represents some snapshots of the flaring active region, taken by SDO/AIA 1600 \AA. 
The SDO/AIA data is available from \url{http://jsoc.stanford.edu/ajax/lookdata.html}.
Some bright points can be seen in panel (a), which shows the start of the flare events. 
Panels (b) - (d) represent the flare ribbons when GOES XRS-B approaches the peak of each event. 
The distance between the ribbons can be estimated as being approximately 32.5 Mm.
We derive the loop length as 51 Mm, regarding the ribbon distance as the diameter of a semicircular loop.
We apply this value to the numerical simulation described in the next subsection. 

\subsection{Numerical simulation}
To solve for the coronal loops heated by the flare, we use the hydrodynamic simulation of the solar flare model given in the CANS (Coordinated Astronomical Numerical Software \footnote[1]{The source code and documents are available at the website \url{http://www.astro.phys.s.chiba-u.ac.jp/netlab/astro/}}) 1D package, developed by T. Yokoyama. 
The simulation set up here is the almost identical to that of \citet{Hori1997} or \citet{Imada2012}.
The fundamental equations are as follows:

\begin{eqnarray}
\frac{\partial}{\partial t}(\rho S) + \frac{\partial}{\partial x}\left(\rho V_x S\right) = 0 \\ 
\frac{\partial}{\partial t}\left(\rho V_x S\right) + \frac{\partial}{\partial x}\left[\left(\rho V_x^2 + p\right)S\right] = \rho g S \\ 
\nonumber
\frac{\partial}{\partial t}\left[\left(\frac{p}{\gamma - 1} + \frac{1}{2}\rho V_x^2\right)S\right] + \\
\nonumber
\frac{\partial}{\partial x}\left[\left(\frac{\gamma}{\gamma - 1}p + \frac{1}{2}\rho V_x^2\right)V_x S - \kappa \frac{\partial T}{\partial x}S\right]= \\
\left(\rho g V_x + H - R + H_f\right)S \\ 
p = \frac{k_B}{m}\rho T
\end{eqnarray}

where, $p$, $T$, $v_x$, $\rho$, $\gamma=5/3$, $S$, $g$, $H$, $R$, $H_f$, $k_B$, $\kappa$, and $m$ represent the pressure, temperature, plasma velocity along the loop, density, heat capacity ratio, cross-sectional area, gravitational acceleration, static heating, radiative cooling, flare heating, Boltzmann constant, thermal conductivity, and mean particle mass, respectively. 
The simulation assumes that the length and cross section of the loop do not change in time, that the cross section is uniform along the loop, that the flow is inviscid and compressible, and that the location where the flare occurs is fixed at the loop top. 
The loop half-length is about 25.5 Mm, as estimated from the SDO/AIA observation. 
The Spitzer thermal conductivity \citep{Spitzer1956}, optically thin radiative cooling, and gravity are taken into account.
An approximation to correct the effects of high-density plasma is included in the radiative cooling model (details are written in the CANS documentation).
This simulation includes not only the corona but also the transition region and chromosphere. 
The flare energy input is represented by the following equations:

\begin{eqnarray}
H_f = H_{f0} \cdot q(t) \cdot f(x) \cdot g(x)\\
q(t) = \frac{1}{4}\left\{ 1 + \tanh{\frac{t}{0.1 \tau_0}}\right\}\left\{1-\tanh{\left(\frac{t-\tau_f}{0.1 \tau_0}\right)}\right\}\\
f(x) = \frac{1}{\sqrt{2 \pi}} \exp{\left[ -\frac{(x-L)^2}{2w_f^2}\right]} \\
g(x) = \frac{1}{2} \left\{ 1 + \tanh{ \left( \frac{x-20 \mathcal{H}_0}{3 \mathcal{H}_0} \right)} \right\}
\end{eqnarray}

where, $\mathcal{H}_0=200~\mathrm{km}$ and $\tau_0=20~\mathrm{s}$ represent the scale height and sound wave traveling time at the surface ($x=0$).
The heating rate, $H_{f0}$, is $120~\mathrm{erg~cm^{-3}~s^{-1}}$ for all observed flares in this paper, which is optimized by comparing observed and estimated XRS-A peak intensities as described in Subsection~\ref{sec:conv}.
$q(t)$ is a function of time to make the heating impulsive. 
The duration and width along the loop of the flare, $\tau_f$ and $w_f$, are fixed at 60 s and 6000 km, respectively. 
The role of $g(x)$ is to prevent the heat pulse from inputting directly into the chromosphere.
Flare heating begins at the start of the simulation.
This simulation uses the modified Lax-Wendroff scheme, which is second-order accurate in both space and time. 

\begin{figure*}
\begin{center}
\includegraphics[width=1\linewidth]{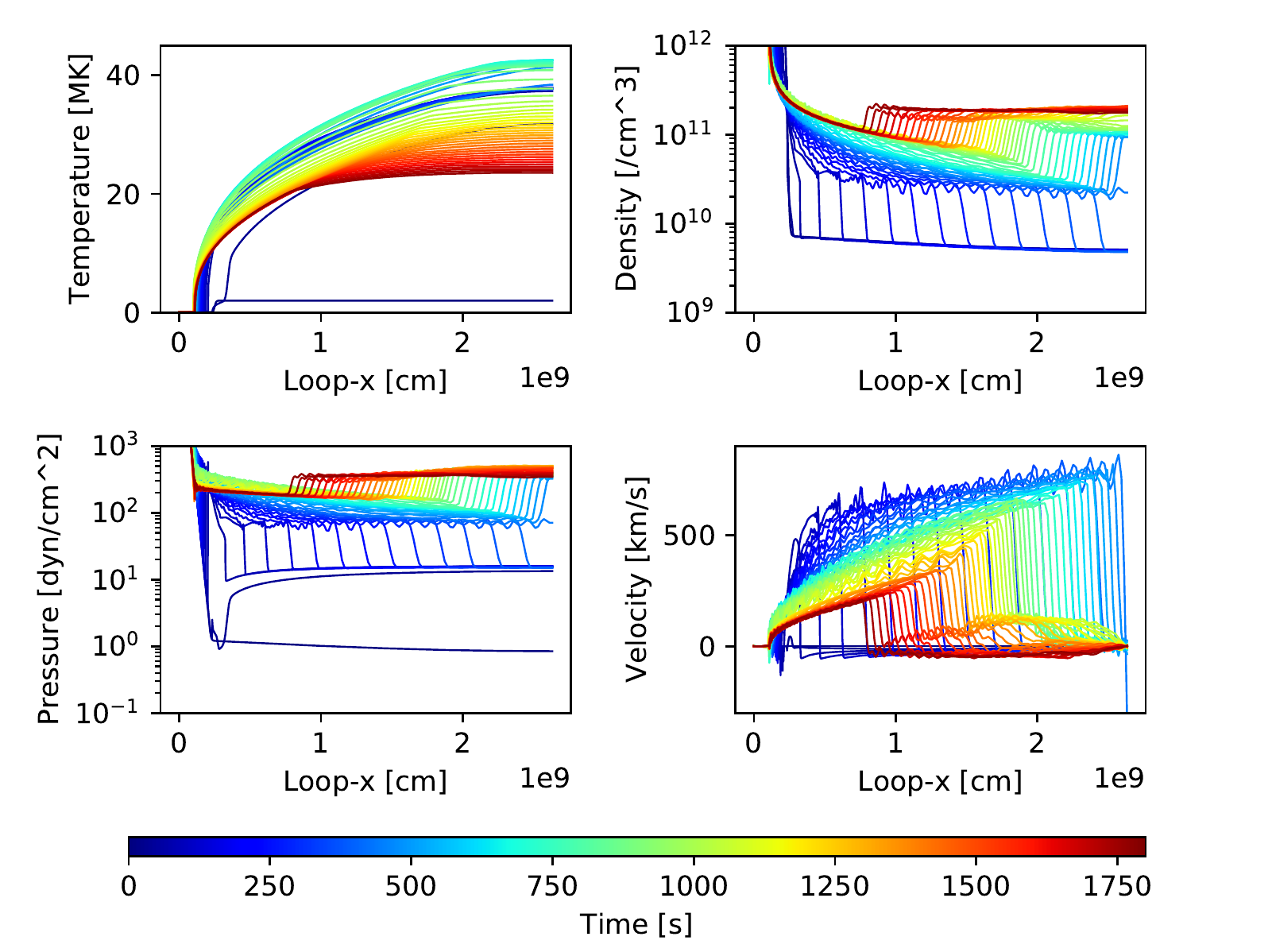}
\caption{Results of the hydrodynamic simulation. Each panel presents the time evolution of the temperature distribution (left-top), density (right-top), pressure (left-bottom), and plasma velocity along the loop (right-bottom) with 30 s cadence. Each horizontal axis represents the loop coordinates from the surface to the loop top. Line color indicates the progress of time (from blue to red).}\label{fig:cans}
\end{center}
\end{figure*}

Figure~\ref{fig:cans} shows the simulation result which is used for all three flares. 
Each panel presents the temporal variation and spatial distribution of temperature, density, pressure, and plasma velocity along the loop, respectively. 
As each horizontal axis indicates coordinates along the loop, the left-hand edge ($x=0$) is the surface and the right-hand edge ($x \simeq 13$ Mm) is the loop top. 
The region where the temperature and pressure change rapidly ($x\simeq 0.3$ Mm) is the transition region, and the chromosphere sits to the left of this.
The line color indicates the progress of time in the simulation, from red to yellow. 
When a flare occurs at the loop top, the temperature there is increased and is transported to the foot points of the loop by thermal conduction. 
Then, the temperature and pressure of the chromosphere is rapidly increased by the incoming high temperature plasma. 
As a result, high-density plasma in the chromosphere is ejected into the corona by the pressure gradient force, this is referred to as chromospheric evaporation.
Consequently, the coronal loop is filled with high-density plasma, and emits soft X-ray and EUV irradiances. 

\subsection{Atomic database}
We obtain the temporal variation of the synthetic X-ray and EUV spectra of the simulated coronal loop, using the atomic database CHIANTI version 8.0 \citep{DelZanna2015}. 
The calculated wavelength range is from 0.5 to 1060.5 \AA~ with a 1 \AA~ resolution. 
To reduce computational cost, we prepare a spectrum table for each temperature range, which runs from $10^{5.0}$ to $10^{7.5}$ K with a 0.1 resolution in the logarithmic scale. 
We obtain the spectrum with the closest-matching temperature for each time and grid of the loop, and multiply this by the ratio between the emission measures of the table and those of the simulation result. 
The emission measure of each grid is assumed to be the density squared. 
We calculate the spectrum of the loop only for regions higher than the transition region, so as to neglect emissions from the transition region and the chromosphere, because we do not calculate radiative transfer in this paper. 
We define the bottom of the corona as the region closest to the surface where the temperature is larger than 0.1 MK for each time. 
The abundance file name is \verb|sun_coronal_2012_schmelz| and the free-free continuum emission is included. 
To simplify the problem, we assume that the intensities of XRS-A and -B are a summation of the calculated spectrum in the wavelength range from 0.5 to 4 \AA~ and from 1 to 8 \AA, respectively.

\begin{figure} 
\centerline{\includegraphics[width=1\linewidth]{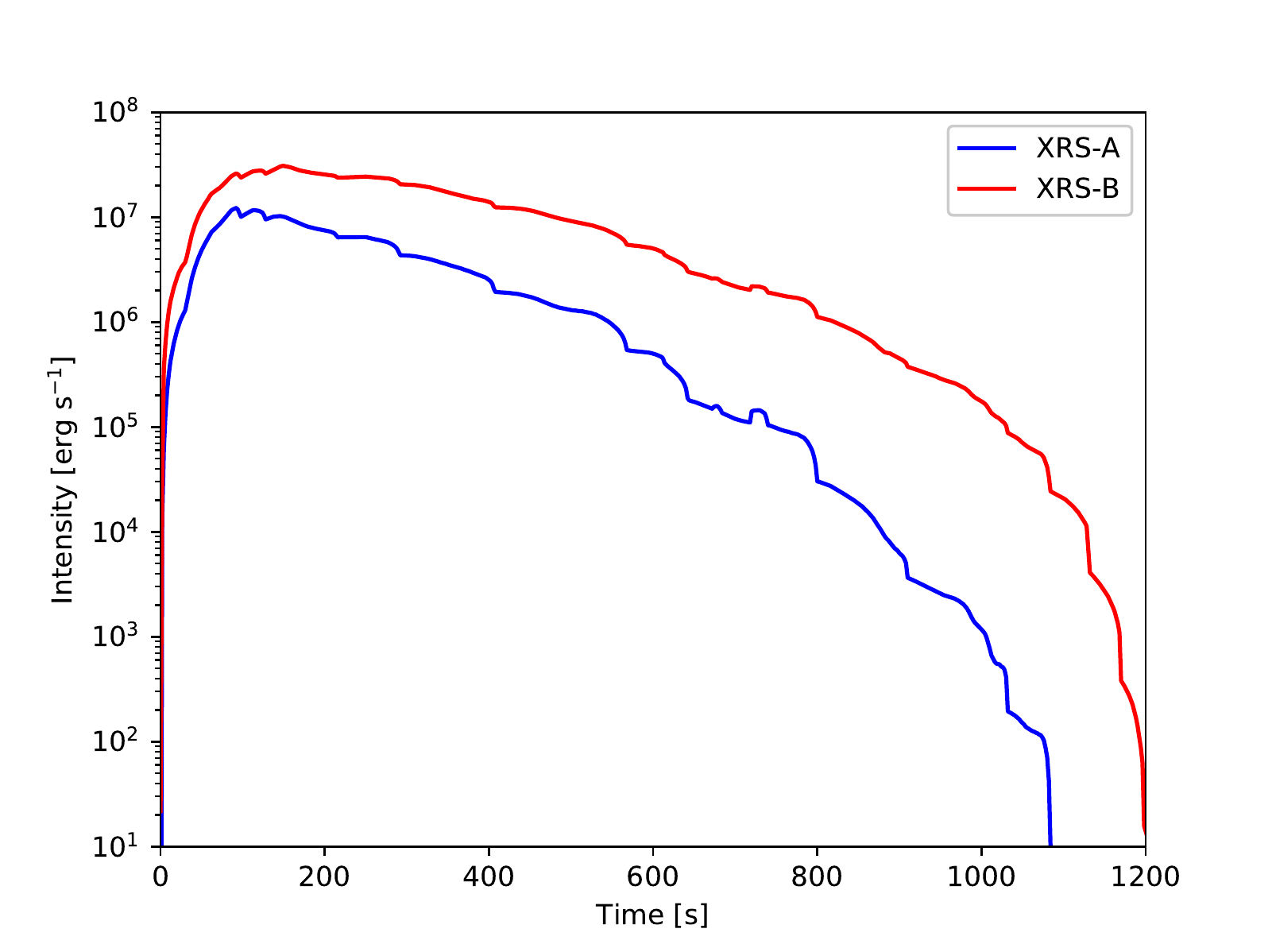}}
\caption{Light curves of XRS-A and -B calculated from the simulation result depicted in Figure~\ref{fig:cans} using the atomic database CHIANTI. The light curve duration we use for the conversion runs from $t=0$ to when the intensity of XRS-B returns to its original value ($t \simeq 1200$).}\label{fig:chianti}
\end{figure}

Figure~\ref{fig:chianti} represents both XRS-A and -B light curves calculated from the simulation of Figure~\ref{fig:cans}. 
As mentioned in the simulation result, light curves are spontaneously increased after the flare occurrence, due to the chromospheric evaporation, and decay with the cooling of the coronal loop. 

\subsection{Conversion from GOES X-ray to EUV spectrum}\label{sec:conv}
We introduce a new method to convert the observed GOES XRS-B light curve to those of other wavelengths, notably EUV irradiances. 
Figure~\ref{fig:chart} represents the flow of the conversion.
The estimated dynamic spectrum can be described as follows.
\begin{equation}
I_\mathrm{estim}(t, \lambda) = R_2 \sum_\tau R_1(\tau)I_\mathrm{sim}(t-\tau, \lambda). \label{eq:conv}
\end{equation}
$R_1$ and $R_2$ are conversion coefficients and $I_\mathrm{sim}$ is the dynamic spectrum of the simulated coronal loop. 
The duration of simulated dynamic spectrum we use for conversion runs from the beginning of the simulation to when the XRS-B intensity returns to its original value ($t \simeq 1200$). 
Before the conversion, we subtract the background XRS-B intensity - that measured at the beginning of the first flare - from the total fitted light curves, to estimate an EUV irradiance only using the flare component. 
This subtraction hardly changes the results, because the background is negligible for peak intensities, however, the conversion for weaker flares can be affected by the background.
The conversion coefficients $R_1$ and $R_2$ can be calculated using the fitted-observed and simulated XRS-B light curves. 
Firstly, $R_1$ is defined as the ratio between the simulated XRS-B light curve peak and the fitted observation in each observed time (Figure~\ref{fig:chart}a), that is
\begin{equation}
R_1 (t) = \frac{I_{\mathrm{obs}}\left(t, \lambda_\mathrm{XRS-B}\right)}{\mathrm{Max}\left[I_{\mathrm{sim}}\left(\tau, \lambda_{\mathrm{XRS-B}}\right)\right]}.
\end{equation}

Secondly, $I_\mathrm{sim}(\tau, \lambda_\mathrm{XRS-B})$ is convolved with $R_1(t)$. 
$R_2$ is defined as the ratio between the maximum value of the convolved light curve and that of the fitted XRS-B observation (Figure~\ref{fig:chart}b), that is,

\begin{equation}
R_2 = \frac{\mathrm{Max}\left[I_{\mathrm{obs}}\left(t, \lambda_{\mathrm{XRS-B}}\right)\right]}{\mathrm{Max}\left[\sum_\tau R_1(\tau)I_\mathrm{sim}(t-\tau, \lambda_\mathrm{XRS-B})\right]}.
\end{equation}

Finally, applying $R_1$ and $R_2$ to equation~\ref{eq:conv}, the dynamic spectrum of the simulation can be converted to that observed during flare events. 

\begin{figure}
\begin{center}
\includegraphics[width=1\linewidth]{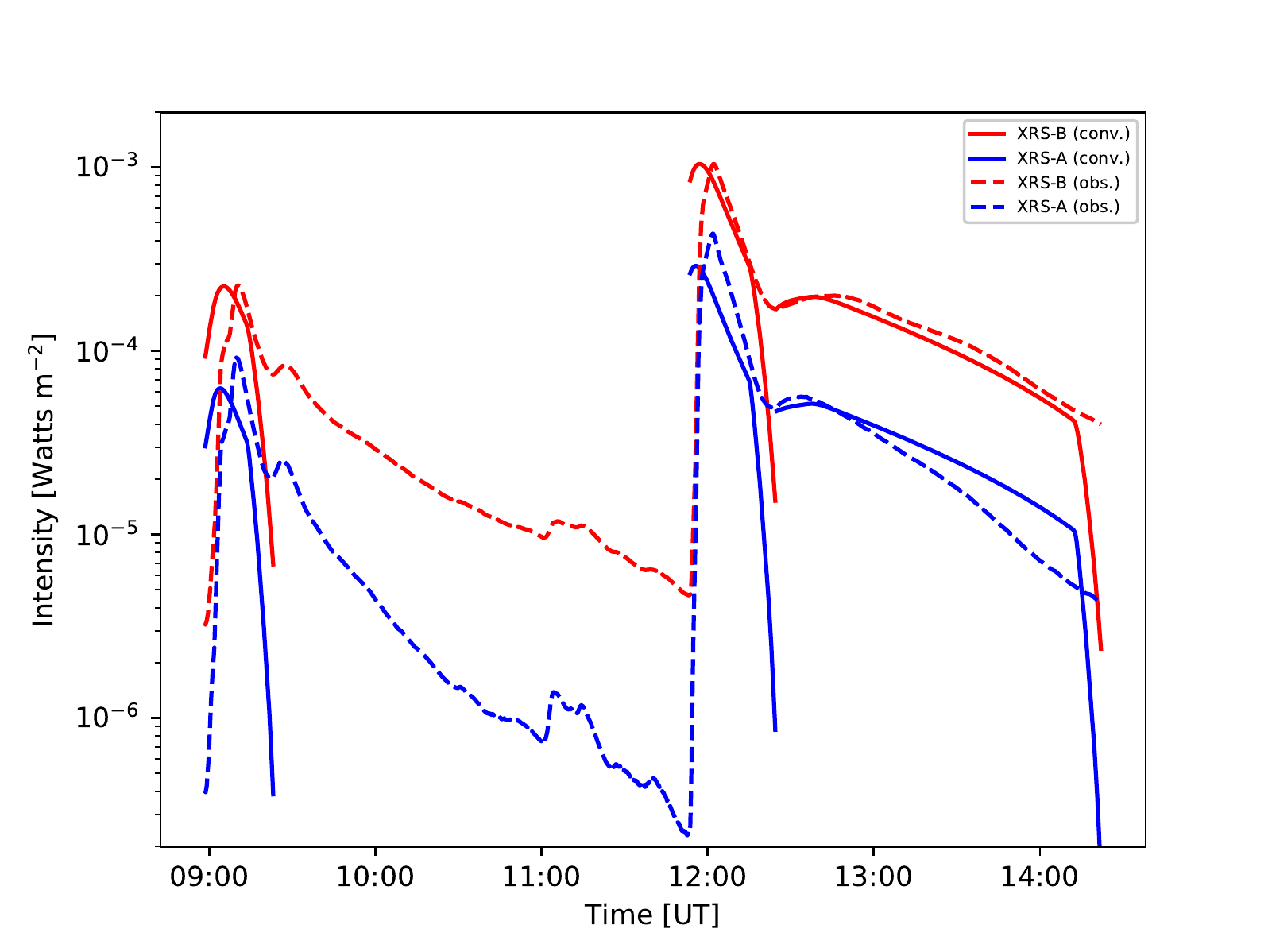}
\caption{Each solid line represents a estimated light curve of GOES XRS during the flares. The dashed red and blue lines represent the observed light curves obtained from XRS-B and -A, respectively.}\label{fig:curve}
\end{center}
\end{figure}

Figure~\ref{fig:curve} represents the observed XRS-A and -B and the converted XRS-A and -B light curves. 
Converted curves are only depicted during each flare event, as mentioned in Table~\ref{tab:flare}. 
We match the peak intensity of the converted XRS-A light curve with that of the observation, by optimizing only the heating rate in the simulation. 
Both light curves plotted across the duration of flare activity seem to be reconstructed well. 

\begin{figure*}
\begin{center}
\includegraphics[width=1\linewidth]{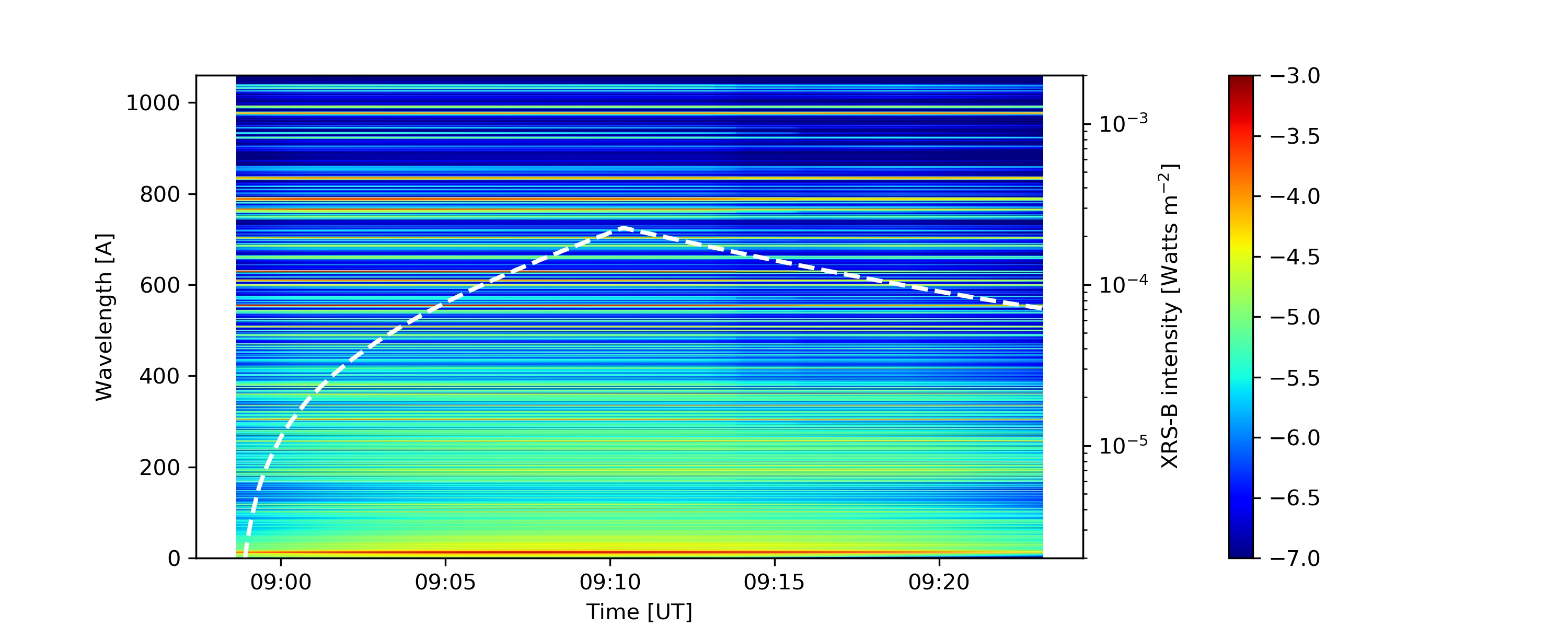}
\includegraphics[width=1\linewidth]{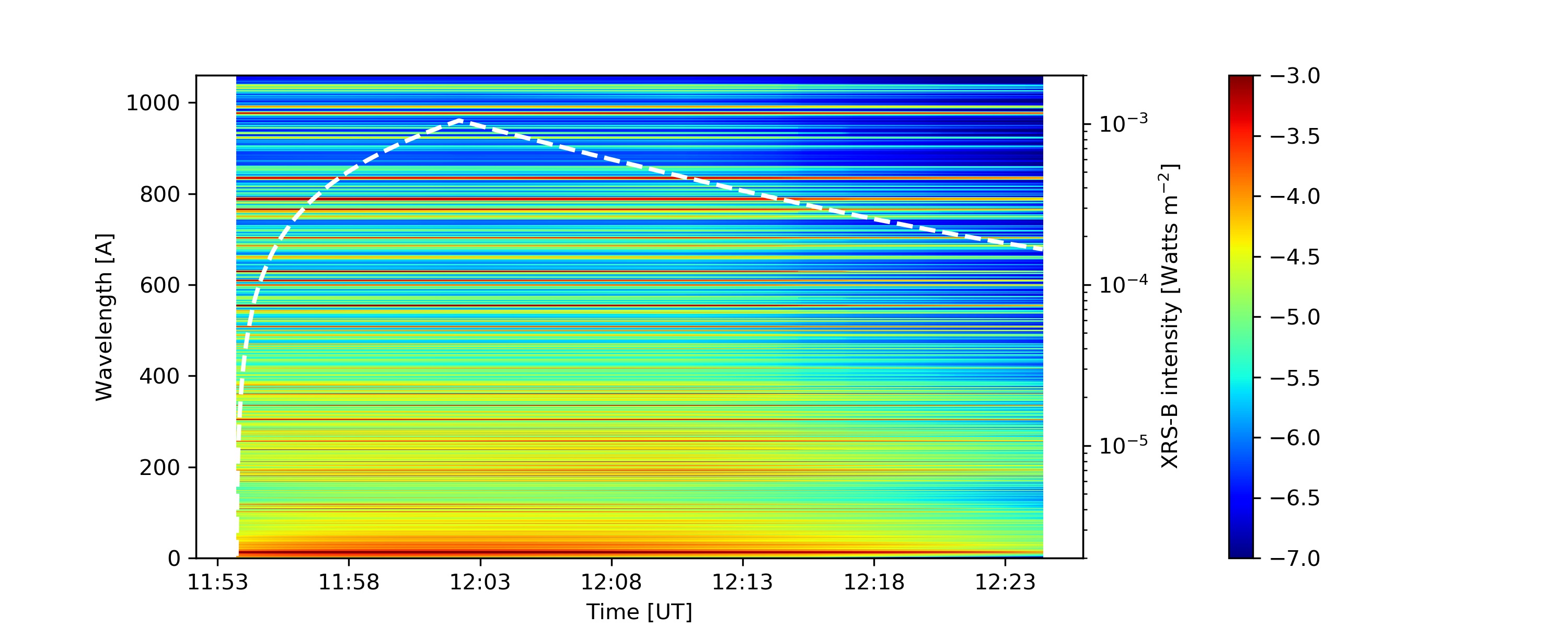}
\includegraphics[width=1\linewidth]{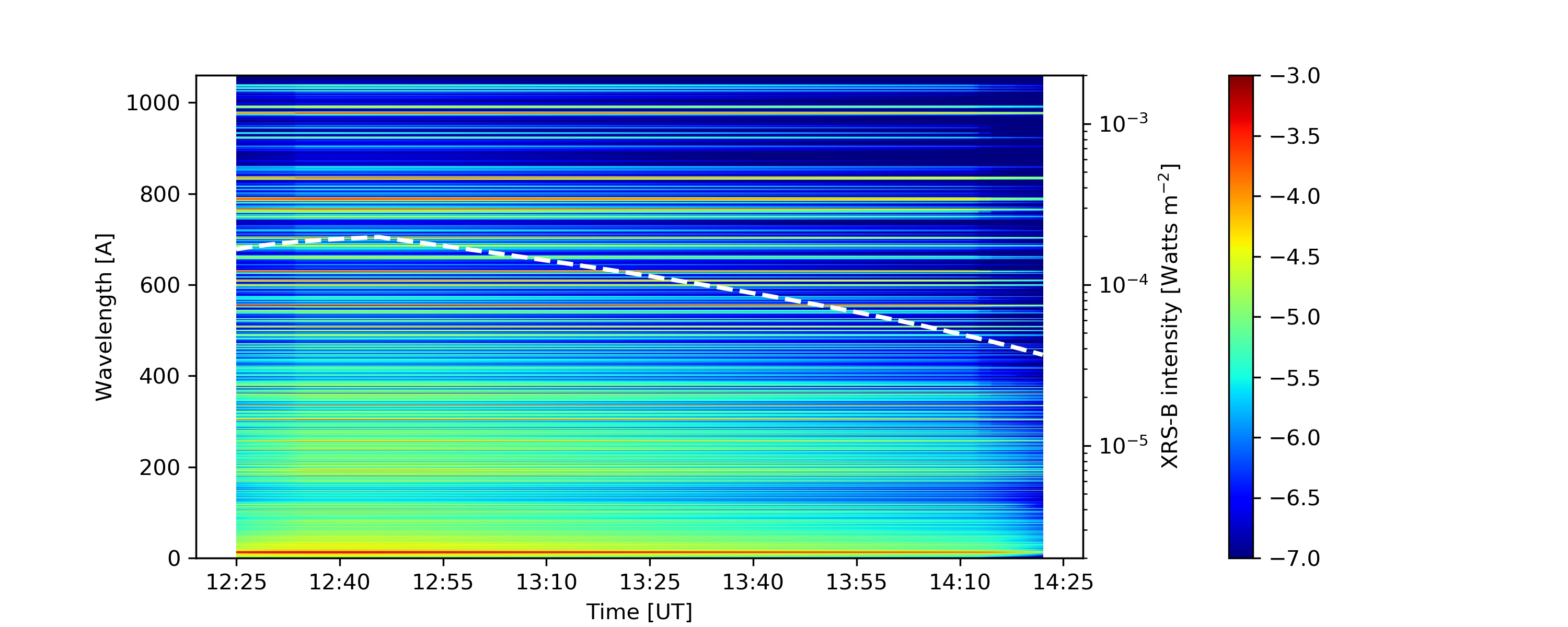}
\caption{Each panel represents the dynamic spectrum during each flare, converted from the observed GOES XRS-B light curves by our method. Color indicates intensity in the logarithmic scale. Each dashed white line represents the fitted GOES XRS-B light curve for each flare.}\label{fig:dyn}
\end{center}
\end{figure*}

Figure~\ref{fig:dyn} represents the converted dynamic spectra during each flare. 
The color indicates the intensity of radiation in the logarithmic scale. 
Each dashed white line represents the fitted GOES XRS-B light curve during flare activity. 
In the next section, we compare the dynamic spectra and some EUV light curves during flares estimated by our method, FISM, and observed by MEGS-B and discuss the validity of our method.

\begin{table}
\begin{center}
  \caption{Optimised $\tau$ of rising and decaying phases of each flare, which reproduce the observed XRS-B light curve well using equation (\ref{eq:tau}). $\pm$ represents the sign in the exponent (divergent or convergent). }
  \begin{tabular}{ccrcr}
 \hline
  	& \multicolumn{2}{c}{Rising phase} & \multicolumn{2}{c}{Decaying phase} \\  \hline
    & $\pm$ & $\tau$ [s] & $\pm$& $\tau$ [s]   \\  \hline
     Flare 1 &+ & 429 & - & 497 \\
     Flare 2 & + & 2757 & - & 565 \\
     Flare 3 & - &  941 & - & 5994 \\
      \hline
  \end{tabular}
  \label{tab:tau}
\end{center}
\end{table}

\section{Discussion and Summary} \label{sec:SD}
We introduced a new method to convert the GOES XRS-B light curves observed during large flare events into EUV dynamic spectra, using the SDO/AIA images, a coronal loop numerical simulation, and an atomic database. 
We apply our method to three large flare events (including an X9.3 class event) as shown in Table~\ref{tab:flare}. 
The estimated dynamic spectra for these events are depicted in Figure~\ref{fig:dyn}. 

\begin{figure*} 
\begin{center}
\includegraphics[width=1\linewidth]{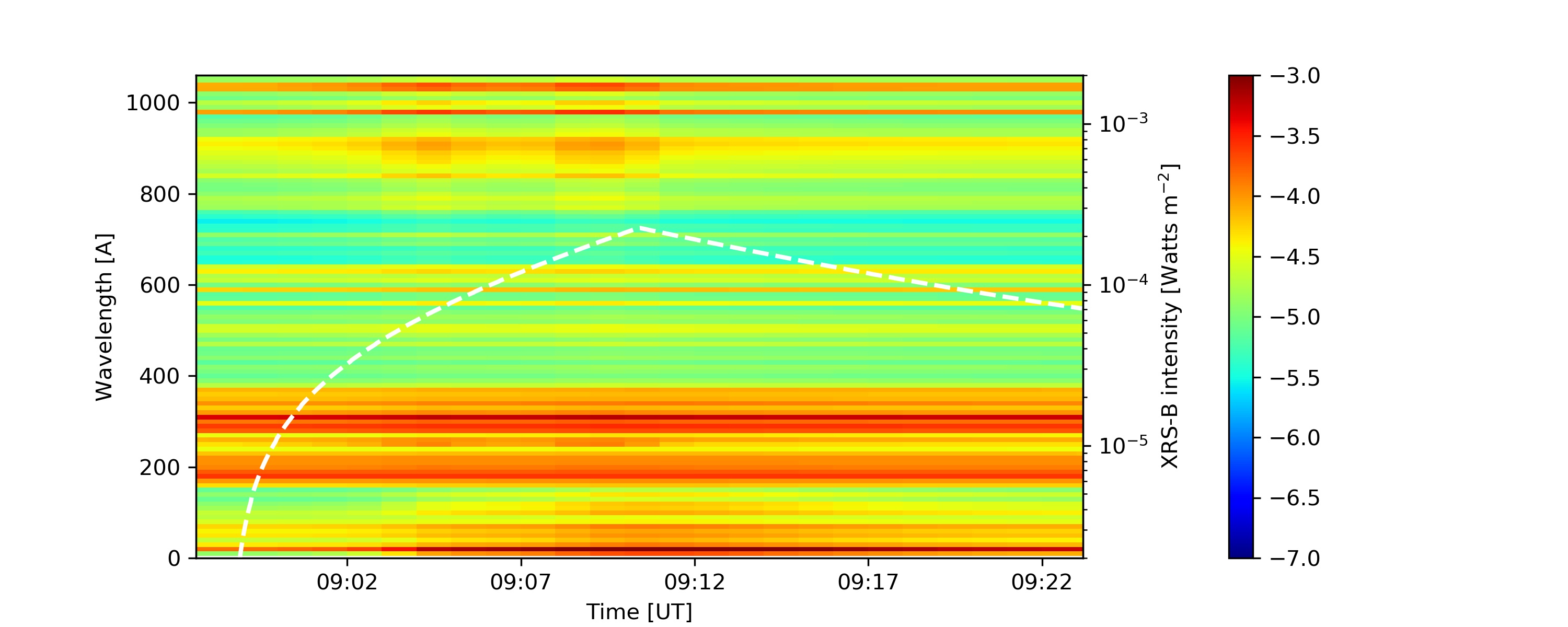}
\includegraphics[width=1\linewidth]{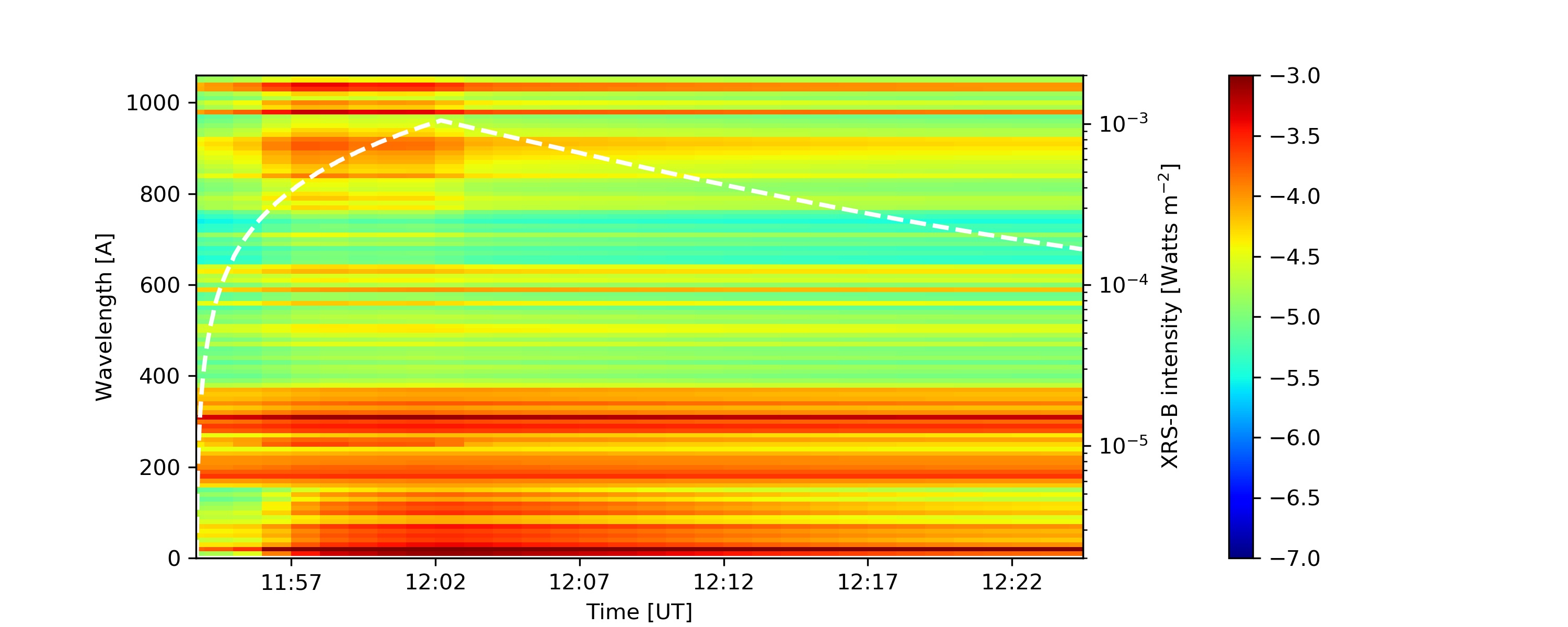}
\includegraphics[width=1\linewidth]{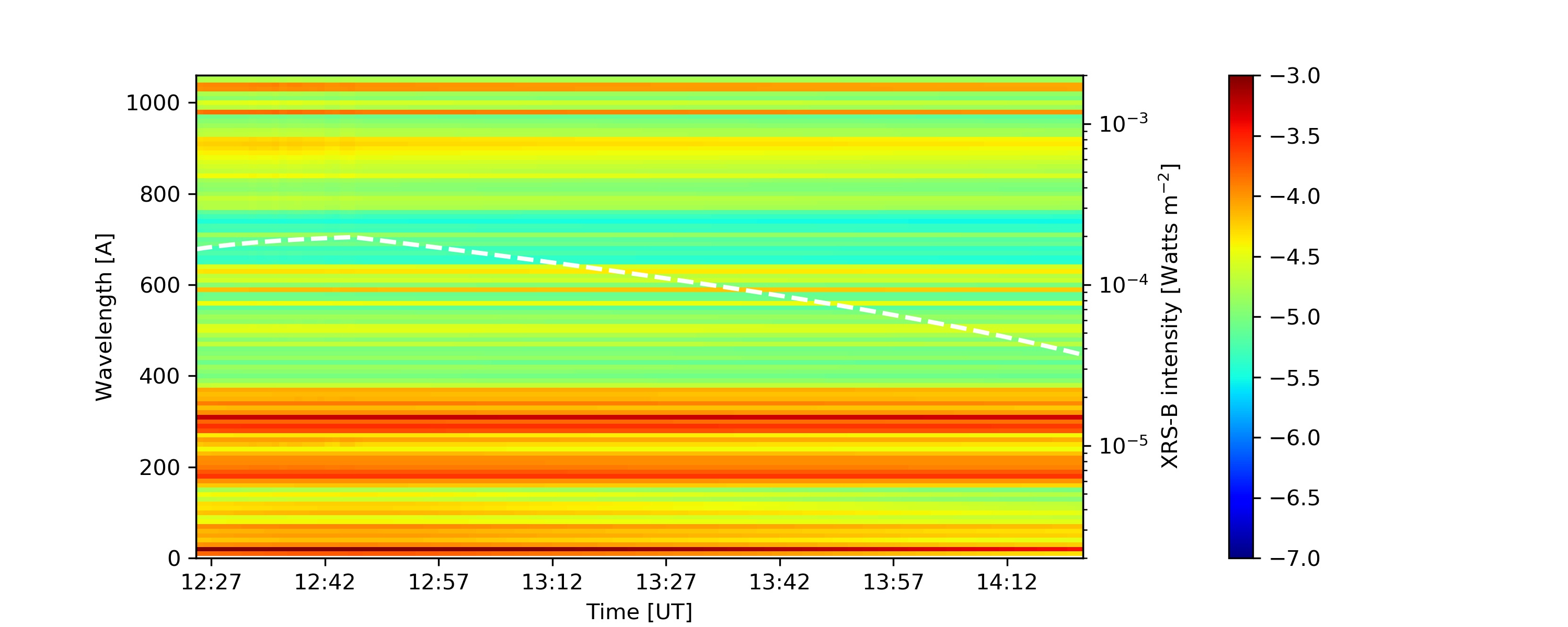}
\caption{Each panel represents the dynamic spectrum created by FISM for each flare. Color indicates intensity in a logarithmic scale. Dashed white lines represent the fitted GOES XRS-B light curve for each flare.}\label{fig:fismw}
\end{center}
\end{figure*}

\begin{figure*} 
\begin{center}
\includegraphics[width=1\linewidth]{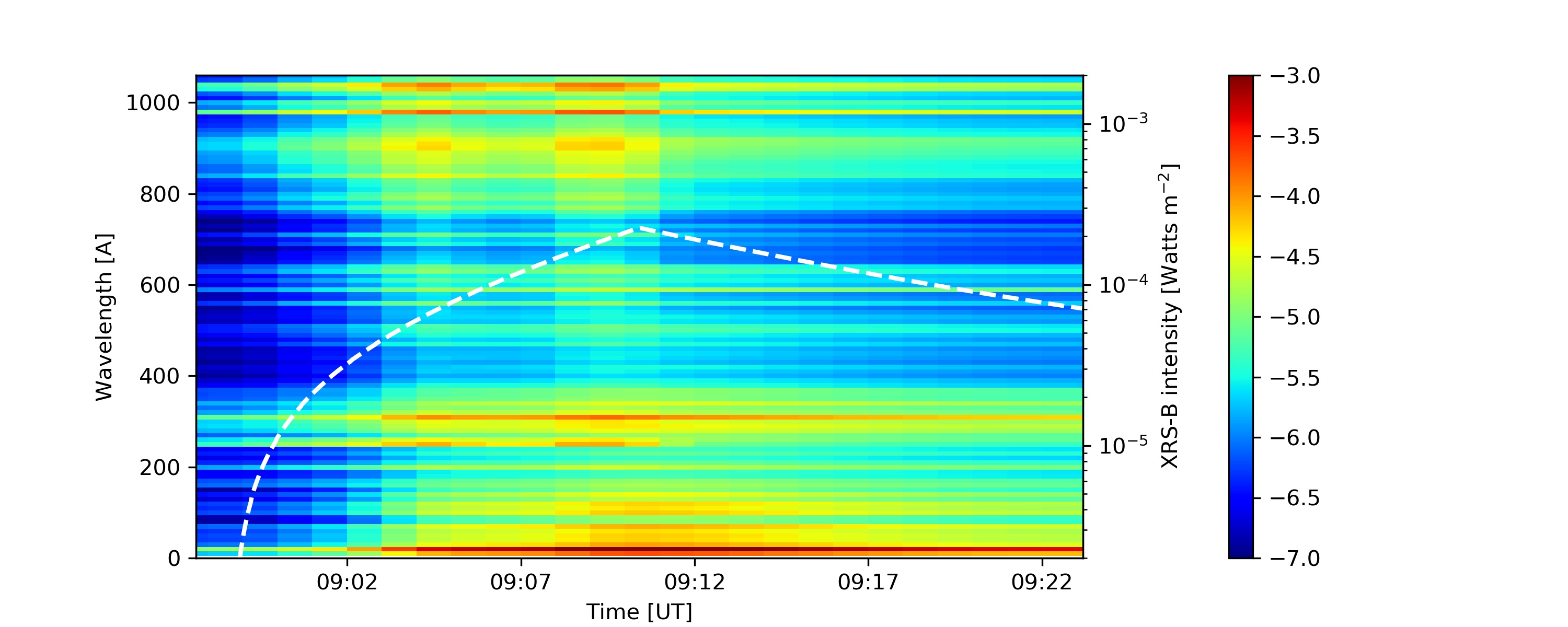}
\includegraphics[width=1\linewidth]{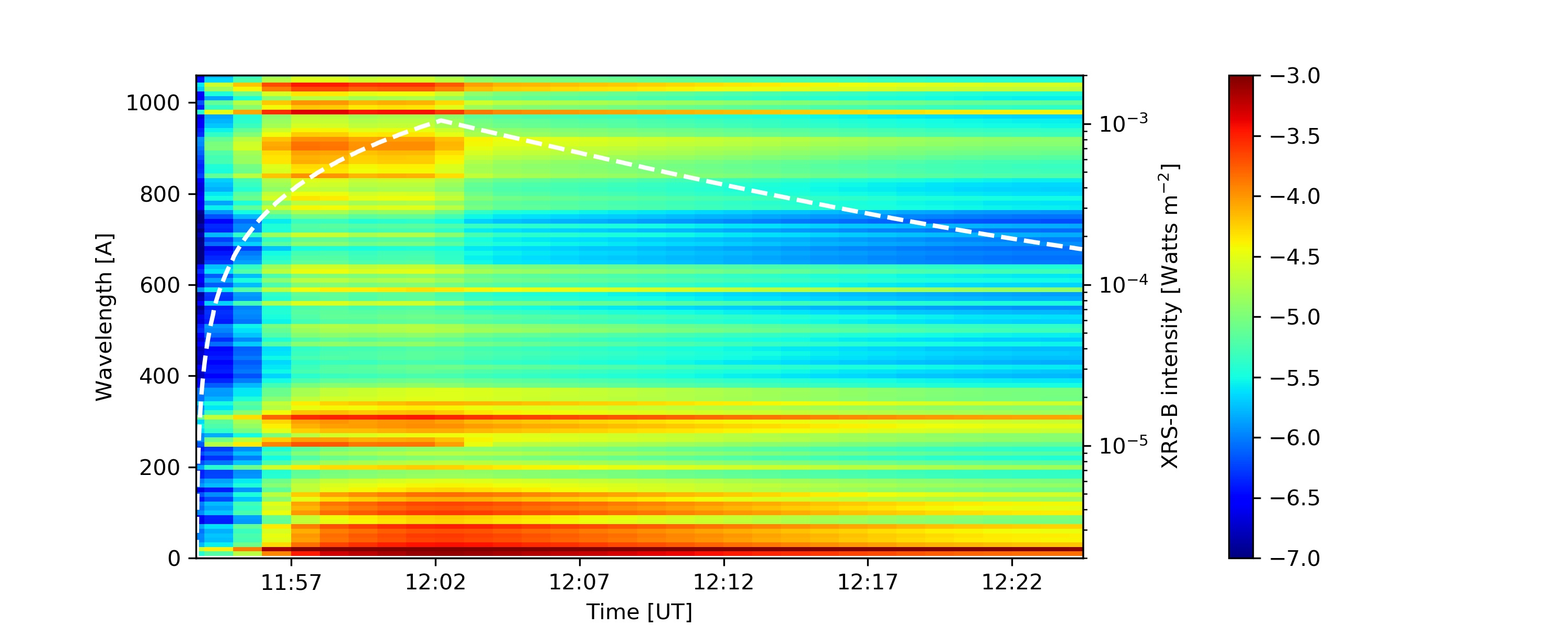}
\includegraphics[width=1\linewidth]{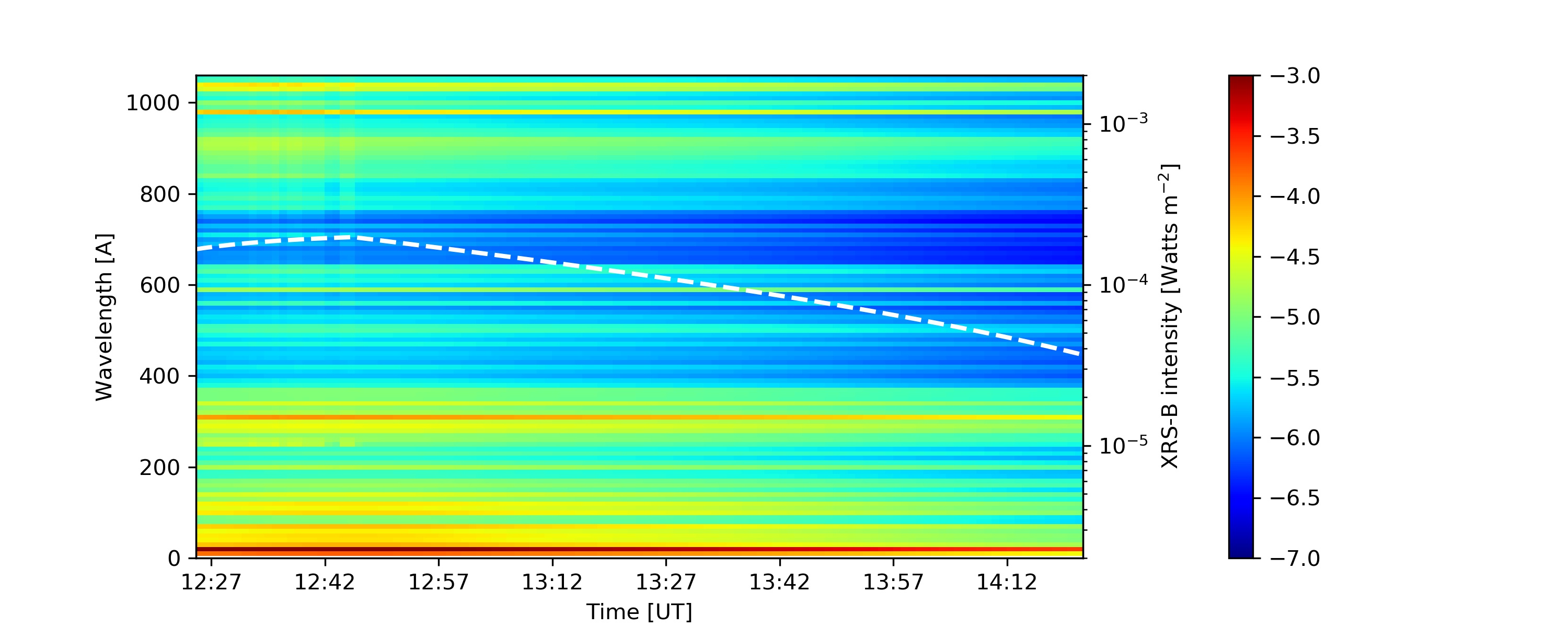}
\caption{Each panel represents the flare component of FISM's dynamic spectrum for each flare, it is made by subtracting the minimum intensity in each wavelength (during each flare) from the original result depicted in Figure~\ref{fig:fismw}.}\label{fig:fismf}
\end{center}
\end{figure*}

Figure~\ref{fig:fismw} shows the dynamic spectra wavelength range from 1 to 1060 \AA~during the flares, created by FISM version 1 which is available from \url{http://lasp.colorado.edu/lisird/data/fism}. 
These results include not only flare components but also daily components such as the changes of irradiance due to the solar cycle and rotation, while our method calculates only flaring coronal loops. 
To compare only flare components, we subtracted the minimum irradiance in each wavelength from the original result (Figure~\ref{fig:fismf}).
Comparing Figure~\ref{fig:dyn} and Figure~\ref{fig:fismf}, it can be seen that the results of our method are similar to those of FISM for wavelengths shorter than approximately 150 \AA.  

\begin{figure*}
\begin{center}
\includegraphics[width=1\linewidth]{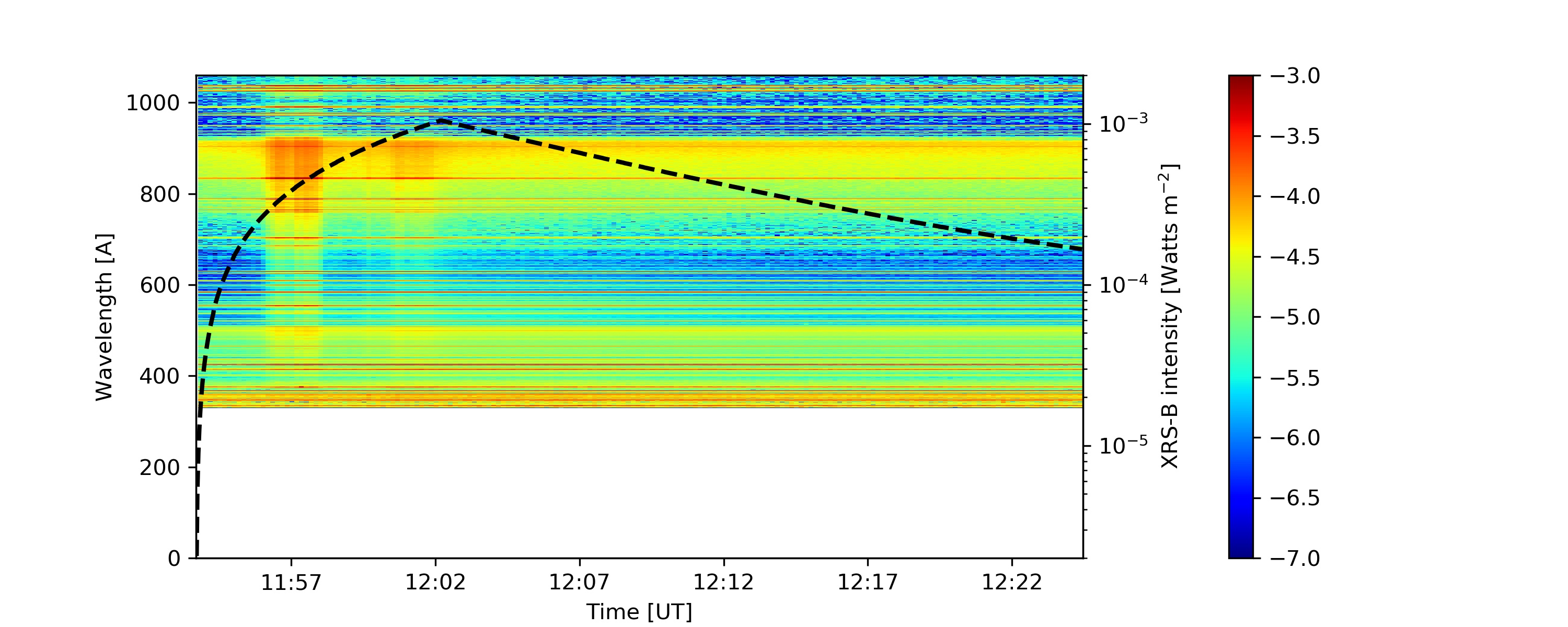}
\includegraphics[width=1\linewidth]{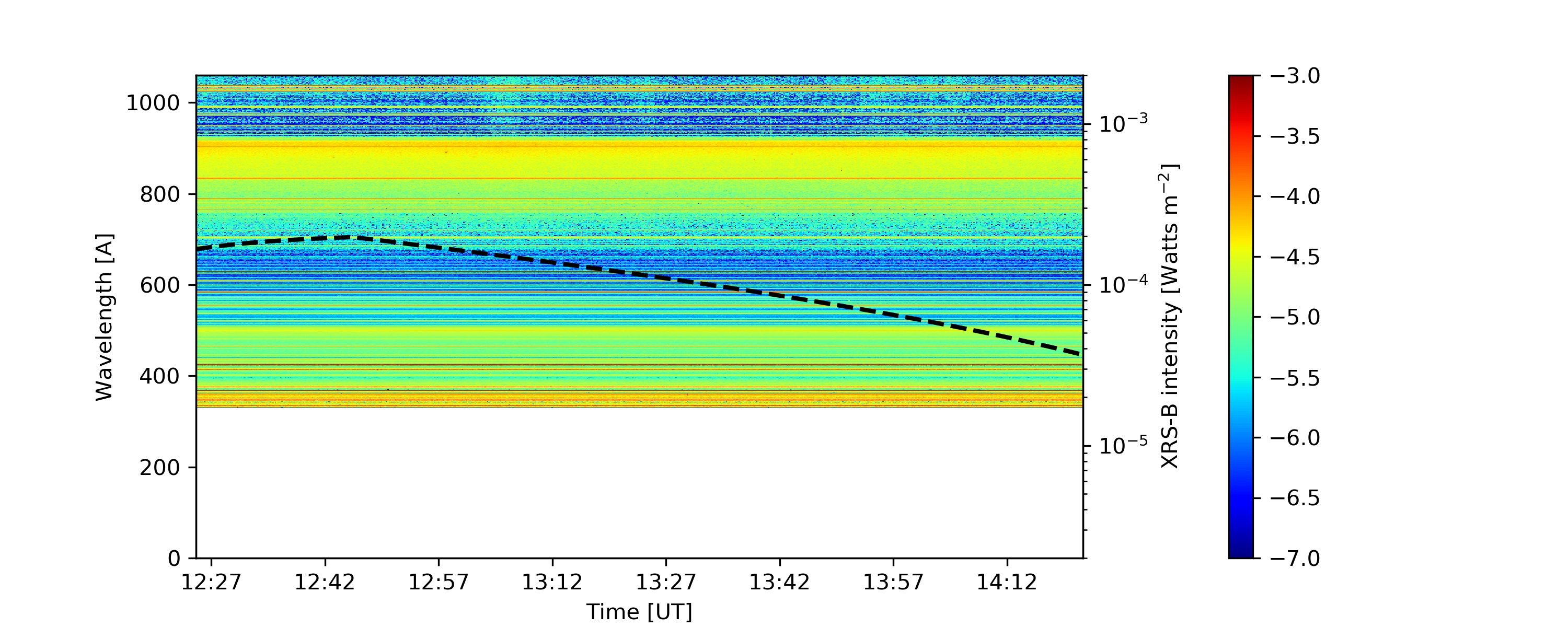}
\caption{Each panel represents the dynamic spectrum for each flare, obtained from SDO/EVE MEGS-B. Color indicates intensity in a logarithmic scale. Dashed black lines represent the fitted GOES XRS-B light curve for each flare. There are no spectral data available for the first flare.}\label{fig:megsw}
\end{center}
\end{figure*}

\begin{figure*}
\begin{center}
\includegraphics[width=1\linewidth]{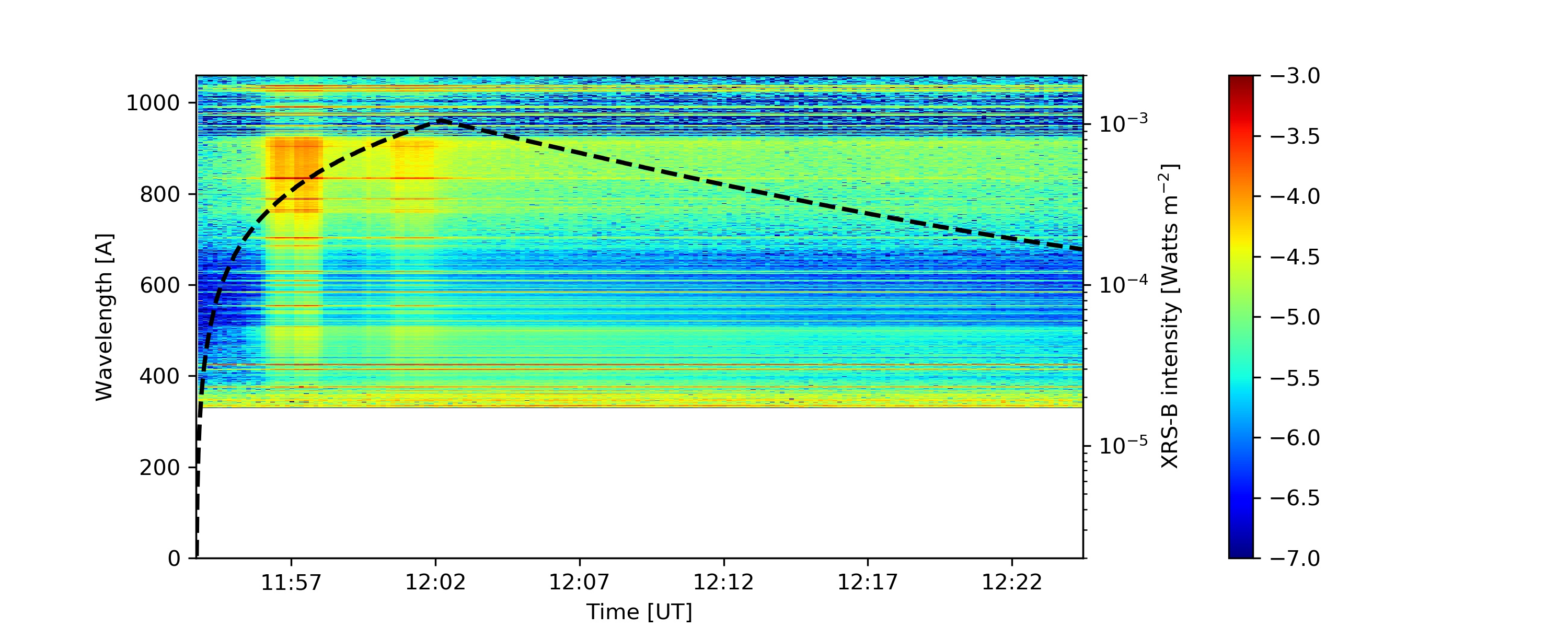}
\includegraphics[width=1\linewidth]{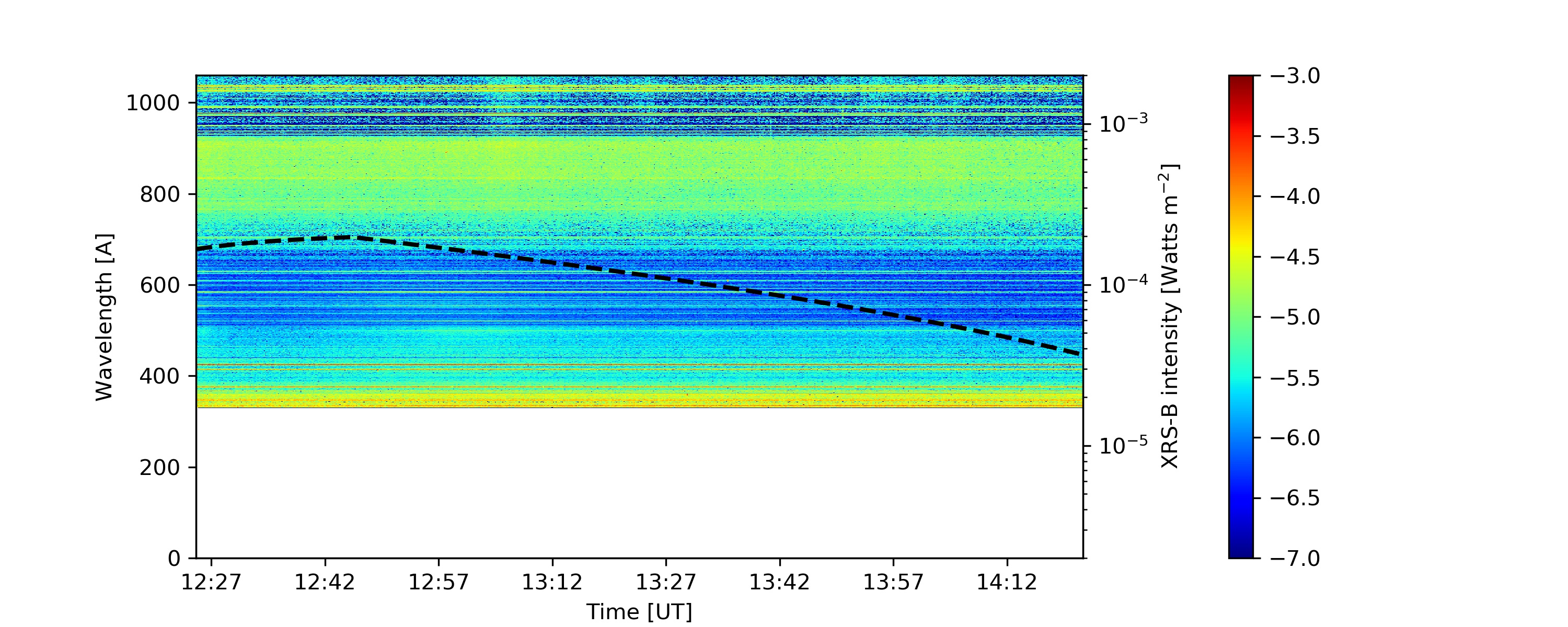}
\caption{Each panel represents the flare component of MEGS-B's dynamic spectrum for each flare, which is made by subtracting the minimum intensity for each wavelength (during each flare) from the original result depicted in Figure~\ref{fig:megsw}.}\label{fig:megsf}
\end{center}
\end{figure*}

Figure~\ref{fig:megsw} represents the dynamic spectra obtained from the MEGS-B observations of the flares, in wavelengths below 1060 \AA, although there are no observations during the first flare. 
We use the EVE level 2 data version 6 which is available from \url{http://lasp.colorado.edu/eve/data_access}.
Between the beginning and peak of the second flare, there is a notable increase in the EUV range. 
According to \cite{Chamberlin2018}, this increase is due to the transition region emissions.
Similarly to the results of FISM, the minimum irradiance for each wavelength are subtracted from the original, so as to compare only flare components (Figure~\ref{fig:megsf}).

\begin{figure*}
\begin{center}
\includegraphics[width=1\linewidth]{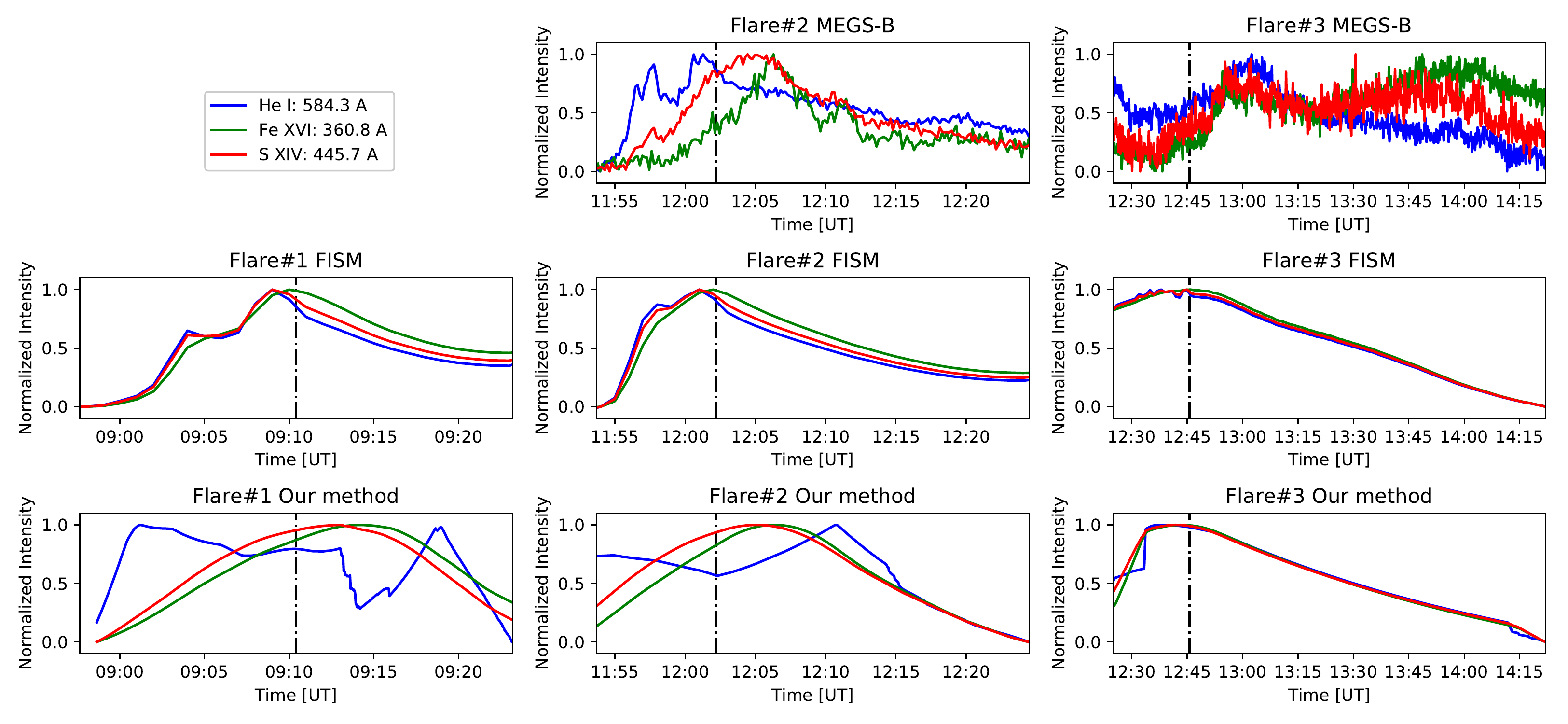}
\caption{Each panel shows the light curves of He I (blue), Fe XVI (green), and S XIV (red) observed by MEGS-B (top), estimated by FISM (middle), and our results (bottom) during each flare. Each black dash-dotted line represents the peak of XRS-B light curve.}\label{fig:plcs}
\end{center}
\end{figure*}

Figure~\ref{fig:plcs} shows the light curves of He I (blue), Fe XVI (green), and S XIV (red) observed by MEGS-B (top), estimated by FISM (middle), and our results (bottom) during each flare. 
Each black dash-dotted line represents the peak of XRS-B observed light curve. 
Each light curve is normalized by its peak value. 
He I 584.3, Fe XVI 360.8, and S XIV 445.7~\AA~are formed around $\log T \simeq 4.5$, $6.8$, and $7.1$, respectively.
The figure shows that we succeeded in reproducing the peak delays of Fe XVI and S XIV lines in the second flare while the FISM cannot reproduce them. 
The peaks are delayed by approximately 15 min from the flare occurrence, which agrees with typical observations \citep{Woods2011, Cheng2019}. 
On the other hand, He I emission has peaks both before and after the soft X-ray peak of the first and second flares. 
The earlier He I peaks are somewhat consistent with impulsive transition region emission as observed by MEGS-B. 
However, the later He I peaks seem to be wrong reproductions due to the lack of calculation of radiative transfer. 
Mostly, In our model, the delay of the peak is defined by the formation temperature of each emission. 
The higher the formation temperature, the smaller the peak delay. 
This tendency can be seen even non-flaring coronal loops probably as a result of the nanoflare heating \citep{viall2012}.

\begin{figure*}
\begin{center}
\includegraphics[width=1\linewidth]{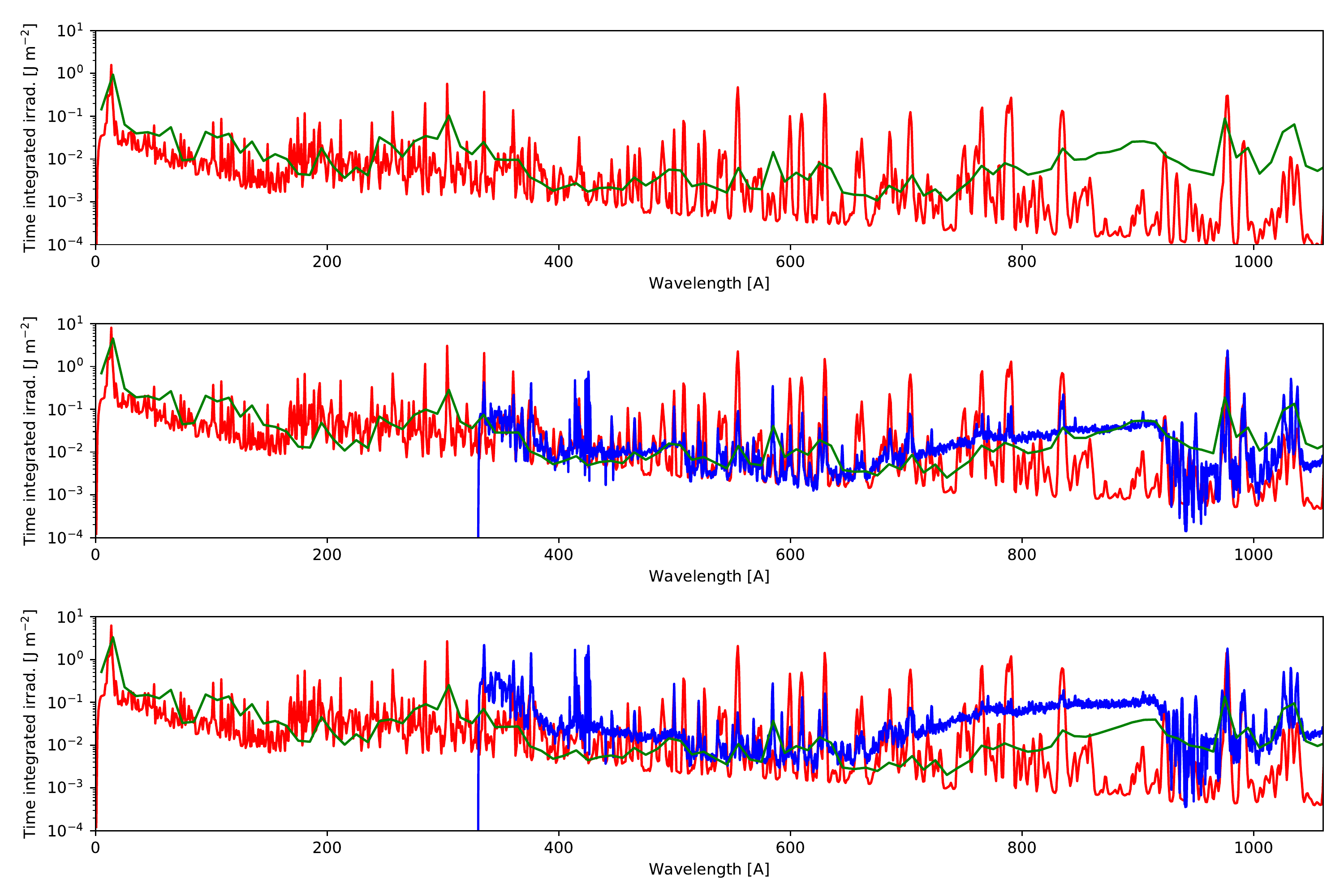}
\caption{Each panel represents the time-integrated irradiance of the flare component for each flare in each wavelength, obtained from our method (red), FISM (green), and MEGS-B observations (blue).}\label{fig:comp}
\end{center}
\end{figure*}

Figure~\ref{fig:comp} represents the time-integrated irradiance of flare components during each flare, for each wavelength.
The solid red, green, and blue lines indicate the values obtained from our method, FISM, and MEGS-B observations, respectively. 
The integration duration runs from the start to the end of each flare described in Table~\ref{tab:flare}. 
Due to the lower wavelength resolution, each line intensity estimated by the FISM is obscured.
Comparing our method with FISM, we can see that the reported trends are similar to each other, however, the intensities of FISM are generally stronger than those of our method especially for longer wavelengths. 
This is most likely because some emissions from below the transition region ({\it e.g.}, the Lyman continuum) are not included in our method. 
In the second flare, our method reproduces well spectrum observed by MEGS-B, with a higher resolution than FISM. 
Some of EUV peaks are estimated stronger by our method than the observation because radiative transfer is not calculated in our method. 
By contrast, in the third flare, the results of both our method and FISM are worse than those of the second flare. 
This is likely because the properties of the real flare lie outside the parameters used in the numerical simulation, and deviate from the statistics of past events due to its occurrence just after the X9.3 flare.

In this paper, we suggested a new method to estimate the soft X-ray and EUV dynamic spectra produced by a solar flare, using a numerical simulation and an atomic database. 
The method reproduces well the observed XRS-A and -B light curves and peak delays of EUV irradiances, however, it is notable that the continuum and line emissions emitted from the transition region and lower are not estimated by our method well. 
This shortcoming need to be modified because the density response of the thermosphere and ionosphere is more sensitive to the emission from the optically thick region than that of optically thin region, which was revealed by comparing two X17 flares occurred at the disc center and the limb \citep{Sutton2006, Qian2010}.
However, the time-integrated dynamic spectra calculated by our method and FISM show similar trends along the wavelength. 
Moreover, our method is able to reproduce the time-integrated observed line peaks in the EUV irradiance, even for the largest flare in cycle 24.

The method we described is mainly based on the physics of the gradual phase of the flares.
However, emissions of some wavelengths in EUV ({\it e.g.,} 304~\AA) are affected by the high energetic electrons accelerated at the impulsive phase of the flares.
Exactly, FISM attempts to reconstruct the light curves during impulsive phase by employing the Neupert Effect \citep{Neupert1968}.
To validate the time series, the physics at the impulsive phase should be considered. 

We chose these events for estimation because the EUV irradiance emitted from them seemed to have effects on Earth, though observations of them were not carried out. 
One of the next steps is to input our results into the simulation of the ionosphere, and compare the results with observed events such as the Dellinger Effects, which cause communication failure due to the increase of the electron density in the ionosphere \citep{Dellinger1937}. 
This step will help validate our method and facilitate nowcasting/forecasting of the effects of the EUV irradiance on Earth. 
The another step is to apply our method to other flare events which were observed by MEGS-A to validate the method at the shorter wavelengths. 

The authors thank H. Iijima, H. Jin, M. Matsumura for fruitful discussions. 
This work was partially supported by the Grant-in-Aid for 17K14401 and 15H05816 and the Program for Leading Graduate Schools, ``PhD Professional: Gateway to Success in Frontier Asia'' by the Ministry of Education, Culture, Sports, Science and Technology. 
A part of this study was carried out using the computational resources of the Center for Integrated Data Science, Institute for Space-Earth Environmental Research, Nagoya University. 

\clearpage

\clearpage

\clearpage

\clearpage

\bibliographystyle{cas-model2-names}
\bibliography{20200501}

\end{document}